\documentclass{eas}
\usepackage{graphicx,xspace}
\newcommand{\Kepler} {\textit{Kepler}\xspace}

\usepackage[dvips]{color}
\newcommand{\blu}{\textcolor{blue} }

\begin{document}

\title{Observations of tides and circularization in red-giant binaries from {\it Kepler} photometry}
%
\author{P.\,G. Beck$^*$}\address{IRFU, CEA, Universit\'e Paris-Saclay, F-91191\,\&\,Universit\'e Paris Diderot, AIM, Sorbonne Paris Cit\'e, CEA, CNRS, F-91191 Gif-sur-Yvette, France 
\\$^*$ presently at: Instituto de Astrofisica de Canarias\,\&\,Univ, La Laguna, Tenerife, Spain}
\author{S. Mathis$^1$}
\author{T. Kallinger}\address{Institut f\"ur Astronomie der Universit\"at Wien, T\"urkenschanzstr. 17, 1180 Wien, Austria}
\author{R.\,A. Garc{\'i}a$^1$}
\author{M.~Benbakoura$^1$}

%
%
\begin{abstract}
Binary stars are places of complex stellar interactions. While all binaries are in principle converging towards a state of circularization, many eccentric systems are found even in advanced stellar phases. In this work we discuss the sample of binaries with a red-giant component, discovered from observations of the NASA \textit{Kepler} space mission. We first discuss which effects and features of tidal interactions are detectable in photometry, spectroscopy and the seismic analysis. In a second step, the sample of binary systems observed with \Kepler, is compared to the well studied sample of Verbunt\,\&\,Phinney (1995, hereafter VP95). We find that this study of circularization of systems hosting evolving red-giant stars with deep convective envelopes is also well applicable to the red-giant binaries in the sample of \Kepler stars.
\end{abstract}
\runningtitle{Beck et al.: The observations of tides in red-giant binaries with \Kepler}
\maketitle
\vspace{-4mm}
\section{\vspace{-2mm}Introduction}
Binary stars are a place of intense interactions which act  between the two components. In his lecture notes,  Jean-Paul Zahn (2013, and references therein) outlines that binary systems are, in a first approximation, a conservative system of angular momentum, defined by the rotation of the stars and their orbital motion. A non-circularized binary system is generally evolving towards an equilibrium state (Hut 1980). Once the system has reached the equilibrium state of minimum kinetic energy (if the orbital momentum is important enough in comparison with the angular momentum of the primary), the binary orbit is circular, the spin axis of the stars are perpendicular to the orbital plane and the stellar rotation of both components is synchronized with the orbital motion. The main process governing the evolution of the orbit and dynamics of the system and the its components are tidal forces. 

Before a system reaches this equilibrium state, tidal interactions sustain kinetic energy in the stellar interior through tidal waves propagating in the binary components. It is converted into heat through various processes (For complete reviews, we refer the reader to  Mathis\,\&\,Remus, 2013 and  Ogilvie 2014).  The  factors governing the efficiency of tidal interactions are the distance between the two components, their radius and their structure. For stars, dominated by convective layers, the main dissipative process is the turbulent viscous friction applied by convection on the so-called equilibrium tide and tidal inertial waves, driven by the Coriolis acceleration (e.g. Zahn 1977, Remus\,et\,al.\,2012, Mathis\,et\,al.\,2016).
 In stars with large radiative zones,  radiative damping is  the dominating process (Zahn 1977). Because these tidal processes redistribute angular momentum between the orbit and the components, the internal stellar dynamics and geometry of the system are modified.  
In addition to the effect tides have on the spins and the eccentricity of the orbit, they also can force the orientation of the orbit to rotate. For example, apsidal motions are introduced in a binary system through asymmetry of the two stars due to the equilibrium tide. While orbital evolution is governed through dissipation of kinetic energy into heat, apsidal motion puts constraints on the internal mass distribution in stars \hbox{(e.g. Zahn 1966, Remus\,et\,al.\,2012).}

Tides in binary and exoplanet systems are subject to intense investigation since several decades (e.g.\ Zahn 1966, VP95, Ogilvie\,\&\,Lin 2007, Remus\,et\,al.\,2012, Mathis 2015).  Until recently, many of these studies were facing the problem that the observations did not well constrain stellar parameters, unless double-lined binary stars (SB2) were studied in eclipsing systems. 
However, the number of stars, well characterized through a combination of space photometry and ground-based high-resolution spectroscopy, allow us to study the distribution of eccentricities and orbital periods and compare the results with previous studies. To date, there are $\sim$40 known binary systems  hosting an oscillating red-giant primary component. While Frandsen\,et\,al.\,(2013) and Gaulme\,et\,al.\,(2014, 2016, hereafter G14 and G16, respectively) reported on eclipsing binaries, Beck\,et\,al.\,(2014, 2018a, hereafter B14 and B17, respectively) focused on eccentric binary systems, \hbox{showing strong tidal interactions at the periastron.}

In Section\,\ref{sec:features}, we discuss the observable features of tidal interactions in red-giant binaries. In Section\,\ref{sec:eccentricities}, we compare the sample of binary stars, observed with the \Kepler satellite to the sample of cluster binaries analyzed by VP95 and repeat their  analysis for the set of \Kepler stars. \hbox{The summary is given in Section\,\ref{sec:conclusions}.}



\vspace{-2mm}
\section{\vspace{-2mm}The effects of the Equilibrium and Dynamical Tides in Red Giants \label{sec:features}}

\begin{figure}
\centering\vspace{-2mm}
\includegraphics[width=\textwidth,height=60mm]{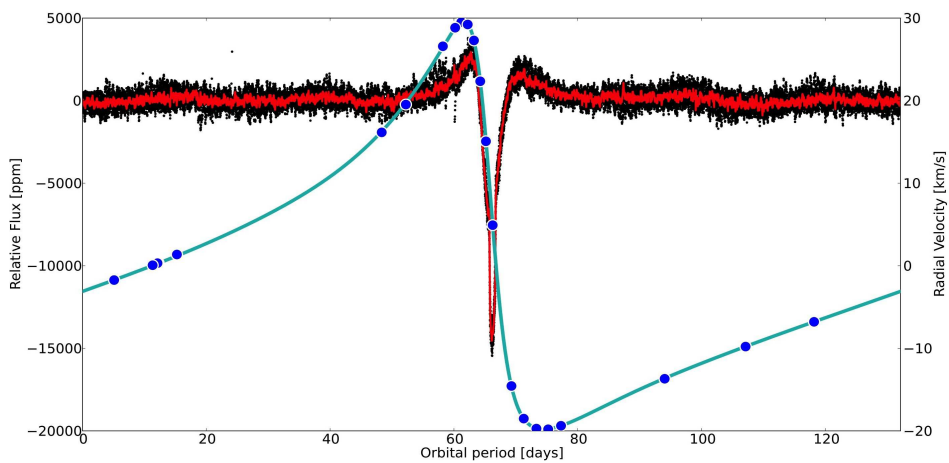}
 \vspace{-2mm}
\caption{
Phase diagram for the eclipsing system KIC\,10614012, showing the effect of the equilibrium tide and stellar eclipses. 
The \Kepler photometry is presented on the left vertical axis. Black data points are the original measurements, calibrated by Garcia\,et\,al.\,(2011) while the red line shows the averaged light curve, which was determined by rebinning the phase diagrams into 30 minutes bins. The radial velocities, \blu{measured with the \textsc{Hermes} spectrograph (Raskin\,et\,al.\,2011) are shown as solid, blue dots whereby the error bars are smaller than the dots} (right vertical axis). The solid blue line represents the orbital model of a Keplerian orbit (Figure from Beck\,2013).
\label{fig:lightcurve}}
\vspace{-4mm}
\end{figure}

The phase diagram shown in Figure\,\ref{fig:lightcurve} depicts the average light and radial-velocity (RV) curve of KIC\,10614012. This  eccentric binary system ($e$\,$\simeq$\,0.71) is a good case to discuss the observational effects of tides in the available observations.



In such eccentric binary systems, the modulation of the brightness due to the distorted stellar shape has been found observationally. This is the signature of the adiabatic component of the hydrostatic equilibrium tide. Due to the  eccentric orbits, the distortion only takes place during the periastron passage, giving rise to temporarily flux variations, that appear with the period of the orbit (see Figure\,\ref{fig:lightcurve}). Because of their particular signature in the light curve which is reminiscent of an echocardiogram, Thompson\,et\,al.\,(2012) coined the term \textit{heartbeat stars} for this class of eccentric binaries. When the binary components are in or close to their periastron, the stars adjust to the tidal force. The surface displacement of each star can be described through the quadrupolar approximation, using three modes with the spherical degree $\ell$=2 and azimuthal order $m$=$\pm$2 and 0. Similar to the effect of rotational splitting of non-radial oscillation modes (see Aerts\,et\,al.\,2010), the individual shape of the lightcurve depends on how an observer sees these three $\ell$=2 modes and therefore on the orientation of the system with respect to the line of sight.
As depicted by the case of KIC\,10614012, numerous configurations of the orbital orientation also allow for the presence of stellar eclipses, superimposed onto the heartbeat event (Fig.\,\ref{fig:lightcurve}). 
The photometric amplitude and duration of the heartbeat event depends on the stellar parameters, the mass ratio as well as eccentricity and separation at periastron. The first models of the shape of these brightness variations were derived by Kumar\,et\,al.\,(1995).

The effect of the dynamical tide, which  typically excites long-periodic inertial waves in convective zones (Ogilvie\,\&\,Li 2007) and gravity waves in radiative layers (Zahn 1975) has not been detected. Such potential tidal oscillations were for example detected in KOI\,54, a heartbeat system containing two A-type stars with external radiative envelopes where tidal gravity waves propagate (Welsh\,et\,al.\,2011). 
For a typical red giant in a binary, the tidal frequency is too small to be able to excite any acoustic mode oscillation through coupling. It would be also questionable if such a mode could actually reach sufficient amplitudes that allow their detection. 
The radiative region in which the radiative damping is the main dissipation mechanism is rather small and located deep under the convective envelope of the giant. Therefore, potential tidal gravity waves may be very difficult to find. Tidal inertial waves, driven by the Coriolis acceleration, can however be excited in the deep convective envelope, if the orbital period is longer than half of the surface rotation period, P$_{\rm orbit}$\,$>$\,0.5$\cdot$P$_{\rm *,rot}$.
Yet, to be detectable in red-giant stars, such modes would need to reach very large oscillation amplitudes in order to compete with the strong signal of the convective granulation. The same frequency region is also contaminated through harmonic frequencies, produced by the Fourier transform of the heartbeat event.
Because spectroscopy is less affected by the granulation noise (e.g.\ Aerts\,et\,al.\,2010) and RV are not sensitive to the photometric variation of the heartbeat event, long and stable RV time series, such as provided by SONG are probably more suited for the search of the dynamical tide in in solar-like oscillators.

Two further effects need to be mentioned, that have not yet been detected in red-giant binaries: apsidal motion and Doppler beaming. For the first one, detection requires substantially longer time bases than already provided by the \Kepler satellite. The latter one is a modulation of the stellar brightness due to the shift of the spectral energy distribution with respect to the filter (Loeb\,\&\,Gaudi 2003). Although it should have substantial contribution to the light curve, it is probably filtered out in the data post processing in typical heartbeat stars (B14).



In a series of papers (Beck 2013, B17) it was tested if tidal interactions in these systems are influencing the seismic properties.
Due to the stochastic nature of solar-like oscillations, it is allowed to close gaps by merging data sets, separated by large gaps. This technique was chosen to analyze KIC\,5006817 to produce a light curve that only contained data during the periastron passage.
In the frequency spectrum no changes of the separation of consecutive radial modes, one of the classical seismic parameters, were found (Beck 2013). This is due to the high-frequency of acoustic waves compared to the orbital frequency. For KIC\,9163796, a different approach was chosen. In a phase diagram, the tidally induced flux modulation were modeled through a high-order Legendre polynome and removed by normalization.  The residual flux does not show any significant peak-to-peak amplitude modulation as function of the orbital phase  (B17).
We therefore conclude that tidal interactions have no immediate effects on acoustic mode parameters.


\vspace{-1mm}
\section{\vspace{-2mm}Eccentricities\,\&\,Circularization of binaries with giant components
\label{sec:eccentricities}}
\begin{figure}
\centering\vspace{-5mm}
\includegraphics[width=\textwidth]{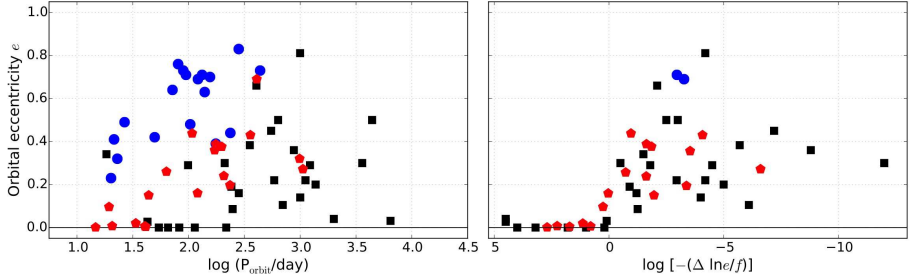}
\caption{
The eccentricity of red-giant binaries in the \Kepler sample and open clusters. Blue dots are marking heartbeat systems (B14, B17), red hexagons indicate binary systems (G14, G16), and black squares mark the systems studied by VP95. The left panel compares the full sample. In the right hand panel, for system for which the mass ratio is known the eccentricity is depicted as function of the expected change of eccentricity $\log [-(\Delta \ln e / f)]$ as defined by VP95. 
\label{fig:eccentricities}}
\vspace{-3mm}
\end{figure}

The number of known binaries is becoming large enough to allow an ensemble analysis. In the left panel of Figure\,\ref{fig:eccentricities}, we compare the period with the eccentricity of the orbit of stars observed by \Kepler, whereby heartbeat stars and eclipses are shown as blue dots and red hexagons, respectively. 

Synchronization of the stellar rotation and alignment of spins will occur long before the circularization of the orbit, because the angular momentum of the slowly rotating stellar envelopes is much smaller than the angular momentum of the binary motion (e.g.\ Zahn 2013). 
Therefore, it is not surprising that the four circularized systems were also found to be synchronized by G14.


In their elaborated analysis of the tidal circularization in binary stars with giant components, VP95 compiled a list of well studied binaries. Because all of these stars are located in open clusters, these stars also have a well constrained mass, radius and age. The set of cluster binaries is compared to the \Kepler sample in Figure\,\ref{fig:eccentricities} as black squares. VP95 pointed out that for these stars a wide spread in period and eccentricities is existing, and that the period alone is not enough to sort the systems. Based on the work of Zahn (1966) on the equilibrium tide they continued to calibrate the formalism for tides and defined the expected eccentricity reduction, $\log [-(\Delta \ln e / f)]$. This quantity depends on the mass of the primary and mass ratio in the system, the radius of the primary and the orbital period of the system. The dimensionless parameter $f$  is determined by the details of the convective turbulent friction and is of order unity. Plotting the eccentricity against the corresponding $\log [-(\Delta \ln e / f)]$ for cluster binaries, depicted as black squares in the right hand panel of Figure\,\ref{fig:eccentricities}, shows that this parameter efficiently separates circularized systems from systems that are still undergoing tidal interaction. 

Although the binary stars in the \Kepler sample are not located in open clusters, we can quantify their mass and radius from asteroseismology (Kallinger\,et\,al. 2010). For those systems, for which also the mass ratio is known, we  calculated $\log [-(\Delta \ln e / f)]$ following the VP95 method and plot them as colored dot in the right-hand panel of Figure\,\ref{fig:eccentricities}. We find the same structure in the diagram as for stars in open clusters and thus confirm the theoretical tidal formalism tested by VP95.\vspace{-2mm}

\newpage
\section{\vspace{-3mm}Discussion\,\&\,Conclusions \label{sec:conclusions}}

We have discussed the observational aspects of tides in binary systems containing red-giant stars. In these systems, the dominating tidal process is the equilibrium tide, leading to characteristic flux modulations at periastron. Because of the stellar structure and related observational challenges the detection of tidal gravity-mode oscillations, is unlikely in red-giant stars. 
We found a good agreement between the circularization  state observed in \Kepler systems and the theoretical predictions, obtained using the equilibrium tide formalism as done for cluster stars by VP95. In a forthcoming  paper, we will extend the analysis and also discuss the effects on synchronization and oscillations, based on \Kepler observations (Beck\,et\,al.\,2018b).\\

{\tiny 
 \textbf{Acknowledgements.} PGB, and RAG  acknowledge the ANR (Agence Nationale de la Recherche, France) program IDEE (n$^\circ$\,ANR-12-BS05-0008) "Interaction Des Etoiles et des Exoplanetes" and   received funding from the CNES grants at CEA. PGB acknowledges the support of the Spanish Ministry of Economy and Competitiveness (MINECO) under the programme 'Juan de la Cierva Incorporacion' (IJCI-2015-26034).
  StM acknowledges support by the ERC through ERC SPIRE grant No.\,647383. The research leading to these results has received funding from the European Community's Seventh Framework Programme ([FP7/2007-2013]) under grant agreement No.\,312844 (SPACEINN) and under grant agreement No.\,269194 (IRSES/ASK).\par }


\end{document}